\documentclass[aps,prd,amsmath,amssymb,superscriptaddress,preprintnumbers,twocolumn,nofootinbib,10pt]{revtex4-1}
\pdfoutput=1
\usepackage{graphicx}
\usepackage{dcolumn}
\usepackage{bm}
\usepackage{amssymb}
\usepackage{latexsym}
\usepackage{booktabs}
\usepackage{amsmath}
\usepackage{multirow}
\usepackage{url}
\usepackage{footnote}
\usepackage{float}
\usepackage{threeparttable}
\usepackage[colorlinks=true, linkcolor=blue, citecolor=blue]{hyperref}
\usepackage[bottom]{footmisc}

\usepackage[normalem]{ulem}
\usepackage{color}
\usepackage{array}
\usepackage{enumerate}
\usepackage{adjustbox}
\usepackage{makecell}
\usepackage{diagbox}
\usepackage{epstopdf}
\usepackage{epsfig}
\usepackage{longtable}
\usepackage{supertabular}
\usepackage{algorithm}
\usepackage{pifont}
\usepackage{algorithmic}
\usepackage{changepage}
\usepackage{setspace}
\usepackage{tabularx}
\newcolumntype{C}{>{\centering\arraybackslash}X}

\begin{document}

\title{New constraints on holographic dark energy with high-redshift galaxy observations from JWST}

\author{Sheng-Han Zhou}
\affiliation{Liaoning Key Laboratory of Cosmology and Astrophysics, College of Sciences, Northeastern University, Shenyang 110819, China}

\author{Tian-Nuo Li}\thanks{Corresponding author}\email{litiannuo@ucas.ac.cn}
\affiliation{School of Fundamental Physics and Mathematical Sciences, Hangzhou Institute for Advanced Study, UCAS, Hangzhou 310024, China}

\author{Guo-Hong Du}
\affiliation{Liaoning Key Laboratory of Cosmology and Astrophysics, College of Sciences, Northeastern University, Shenyang 110819, China}

\author{Yi-Min Zhang}
\affiliation{Liaoning Key Laboratory of Cosmology and Astrophysics, College of Sciences, Northeastern University, Shenyang 110819, China}

\author{Yun-He Li}
\affiliation{Liaoning Key Laboratory of Cosmology and Astrophysics, College of Sciences, Northeastern University, Shenyang 110819, China}

\author{\\Jing-Fei Zhang}
\affiliation{Liaoning Key Laboratory of Cosmology and Astrophysics, College of Sciences, Northeastern University, Shenyang 110819, China}

\author{Xin Zhang}\thanks{Corresponding author}\email{zhangxin@mail.neu.edu.cn}
\affiliation{Liaoning Key Laboratory of Cosmology and Astrophysics, College of Sciences, Northeastern University, Shenyang 110819, China}
\affiliation{MOE Key Laboratory of Data Analytics and Optimization for Smart Industry, Northeastern University, Shenyang 110819, China}
\affiliation{National Frontiers Science Center for Industrial Intelligence and Systems Optimization, Northeastern University, Shenyang 110819, China}

\begin{abstract}
Observations of high-redshift galaxies from the James Webb Space Telescope (JWST) provide a complementary probe of early structure formation. Meanwhile, DESI results have provided further indications of possible dynamical dark energy, prompting investigations of theoretically motivated dark energy models. In this work, we investigate holographic dark energy (HDE) and interacting holographic dark energy (IHDE) using JWST ultraviolet luminosity function measurements, cosmic microwave background (CMB) observations from Planck and ACT, DESI DR2 baryon acoustic oscillation measurements, and DES-Dovekie type~Ia supernova data. We find that JWST has little impact on the parameter constraints of $\Lambda$CDM and HDE but produces more pronounced changes in those of IHDE. For CMB+DESI+DES-Dovekie, adding JWST shifts the holographic parameter $c$ from $1.270^{+0.146}_{-0.096}$ to $1.091\pm0.095$ and the interaction parameter $\beta$ from $0.650^{+0.104}_{-0.034}$ to $0.498^{+0.090}_{-0.061}$, accompanied by an increase in $\Omega_{\rm m}$ and a decrease in $S_8$. For the full data combination, $\chi^2_{\rm JWST}$ decreases by approximately $7$ for both HDE and IHDE compared with $\Lambda$CDM, indicating improved fits to the high-redshift galaxy abundances. Our results highlight the complementary role of high-redshift galaxy observations in probing the HDE and IHDE models.
\end{abstract}

\maketitle

\section{Introduction}\label{sec1}

The $\Lambda$ cold dark matter ($\Lambda$CDM) model consistently describes a broad range of cosmological observations, including the cosmic microwave background (CMB), baryon acoustic oscillations (BAO), and type~Ia supernova (SN)~\cite{Planck:2018vyg,eBOSS:2020yzd,Brout:2022vxf}. However, the ``fine-tuning'' and ``cosmic-coincidence'' problems remain unresolved~\cite{Sahni:1999gb,Bean:2005ru}, while different determinations of the present-day cosmic expansion rate $H_0$ continue to be in tension~\cite{Riess:2021jrx}. This tension and its possible theoretical explanations have been extensively discussed in the literature~\cite{Zhao:2017urm,Guo:2018ans,Poulin:2018cxd,Verde:2019ivm,DiValentino:2019ffd,Vagnozzi:2019ezj,Riess:2019qba,DiValentino:2020zio,Smith:2020rxx,Vagnozzi:2021gjh,Vagnozzi:2021tjv,Gao:2021xnk,Riess:2021jrx,Schoneberg:2021qvd,DiValentino:2021izs,Abdalla:2022yfr,DiValentino:2022fjm,Kamionkowski:2022pkx,Gao:2022ahg,Vagnozzi:2023nrq,Giare:2023xoc,Hu:2023jqc,Poulin:2023lkg,Pedrotti:2025ccw,Cai:2026swf,Schoneberg:2026eys,Tiwari:2026pzk,Schoneberg:2026vaf,Bella:2026zuk,Sabogal:2026ipu,Pedrotti:2026dwj,Liu:2025evk}. The recent Data Release~2 (DR2) of the Dark Energy Spectroscopic Instrument (DESI) has provided more precise measurements of the late-time expansion history, posing further challenges to $\Lambda$CDM. When combined with CMB and SN data, the DESI DR2 results yield a $2.8\sigma$--$4.2\sigma$ preference for the $w_0w_a$CDM model over $\Lambda$CDM, depending on the SN sample adopted~\cite{DESI:2025zgx,DES:2025sig}. The DESI results have attracted considerable attention and prompted extensive discussion in the literature~\cite{Gialamas:2024lyw,Wang:2024dka,Ye:2024ywg,Liu:2024yib,Li:2024qso,Giare:2024gpk,Reboucas:2024smm,Park:2024pew,Jiang:2024xnu,Yang:2024kdo,Li:2025vuh,Wolf:2025jed,Shajib:2025tpd,Chaussidon:2025npr,Pang:2025lvh,RoyChoudhury:2025dhe,Paliathanasis:2025cuc,Scherer:2025esj,Giare:2024oil,Liu:2025mub,Teixeira:2025czm,Specogna:2025guo,Cheng:2025lod,Herold:2025hkb,Ozulker:2025ehg,Ormondroyd:2025iaf,Silva:2025twg,Ishak:2025cay,Fazzari:2025lzd,Zhang:2025lam,Song:2025bio,Li:2025ops,Lee:2025pzo,Santos:2025wiv,Khoury:2025txd,Li:2025ula,Kessler:2025kju,Smith:2025icl,Li:2025owk,Li:2025dwz,Li:2026hwq,Li:2026xaz,Giare:2024smz,Lemos:2025bgy,You:2025uon,Turner:2026vzp,Ribeiro:2026lyd,Wang:2026wrk,Wang:2025znm,Yang:2025gaz,Wang:2026vqw,Yang:2026qmz,Pan:2025qwy,Yang:2026yaq,Du:2025xes,Pedrotti:2026ncd,Yao:2025twv,Mukherjee:2026lnb,Yang:2026vti,Fortunato:2025qxc,Zhou:2026dti,Kulhkampa:2026bfg,Xu:2026sbw,Zhang:2026axl,Ling:2025lmw,Jiang:2026ngo,Liu:2025myr,Li:2025htp,Li:2025muv,Wu:2025wyk,Wu:2024faw,Ling:2026awm}; for recent reviews, see, e.g., Refs.~\cite{CosmoVerseNetwork:2025alb,Giare:2025pzu,Qiu:2025oop,Cai:2025mas,Anchordoqui:2026hys,Li:2026asg,Giare:2026oti,Ren:2026jyw}.

Although phenomenological equation-of-state (EoS) parameterizations, such as the $w_0w_a$CDM model, can accommodate the evolving dark energy (DE) behavior favored by some joint analyses incorporating DESI data, their prescribed functional forms alone cannot identify the physical mechanism responsible for this evolution. Beyond such phenomenological descriptions, a broad range of theoretically motivated DE models has been extensively studied~\cite{Amendola:1999qq,Zhang:2004gc,Cai:2004dk,Zhang:2005rg,Zhang:2005kj,Zhang:2005rj,He:2008si,Majerotto:2009zz,Zhang:2009un,Clemson:2011an,Salvatelli:2013wra,Li:2014cee,Li:2015vla,Wang:2016lxa,DiValentino:2017iww,Pan:2020zza,DiValentino:2020vnx,Nunes:2021zzi,Jin:2021pcv,Guo:2018gyo,Feng:2019jqa,Feng:2017usu,Pan:2023mie,Zhai:2023yny,Hoerning:2023hks,Escamilla:2023shf,Castello:2023zjr,Li:2023gtu,Mishra:2023ueo,Wang:2024vmw,Giare:2024ytc,Halder:2024uao,Zhang:2025dwu,Silva:2025hxw,Wu:2025vrl,vanderWesthuizen:2025rip,Lyu:2025nsd,Sabogal:2026bsw,Li:2026jgg}. One theoretically motivated framework is holographic dark energy (HDE), which is inspired by the ultraviolet--infrared connection associated with the holographic principle. With the future event horizon adopted as the infrared cutoff, the HDE energy density is linked to this horizon scale, giving rise to dynamical evolution~\cite{Li:2004rb,Huang:2004ai,Wang:2004nqa,Zhang:2005hs,Zhang:2005yz,Zhang:2006qu,Ma:2007av,Zhang:2007sh,Li:2013dha,Wang:2016och,Fu:2011ab}. In its standard formulation, HDE assumes that DE and cold dark matter (CDM) are separately conserved. If energy exchange between DE and CDM is allowed, the evolution of their respective energy densities is modified, leading to interacting holographic dark energy (IHDE) models~\cite{Pavon:2005yx,Wang:2005jx,Zhang:2007uh,Zhang:2012uu,Feng:2018yew}.

Recent studies have tested HDE and IHDE using various cosmological observations, including DESI BAO, CMB, and SN data~\cite{Li:2024qus,Li:2024bwr,Shen:2025cjm,Wu:2025vfs,Li:2025vqt,Shen:2026ynj,Huang:2026szy}. In particular,~\citet{Li:2024qus} revisited the HDE and IHDE models using CMB, DESI DR1 BAO, and several SN samples, finding that dark sector interactions favor larger values of the holographic parameter $c$ and reduce the likelihood of phantom-divide crossing compared with HDE. Although these observations provide stringent constraints on HDE and IHDE, the CMB is primarily sensitive to early universe physics, whereas BAO and SN mainly probe the background expansion history. However, the dynamical evolution of HDE and its possible interactions with CDM can also affect the growth of matter perturbations and hence galaxy formation.

The James Webb Space Telescope (JWST) provides observations of high-redshift galaxy abundances, offering a complementary probe for testing HDE and IHDE. With the rapidly growing sample of galaxies at $z\gtrsim7$, JWST measurements have increasingly been incorporated into cosmological analyses, opening a new observational window on the early universe~\cite{Adil:2023ara,Forconi:2023hsj,Comini:2026nsj,Liu:2024yan,Jiang:2024tll,Du:2026qtq,Zhou:2025nkb,Matteri:2025klg,Li:2025jxq,Ellis:2025dpw,Das:2025bnr,Barron:2025dys,Blamart:2025szc,Fairbairn:2026dva,Co:2026klr}. The galaxy ultraviolet luminosity function (UV LF) describes the comoving number density of galaxies per unit UV magnitude, characterizing their abundance as a function of luminosity and redshift~\cite{Bouwens:2022gqg,Donnan2022,Finkelstein2023,Leethochawalit2023}. Through the connection between galaxy UV luminosity and host dark matter halo mass, the UV LF is sensitive to the underlying halo mass function and hence to the growth of matter perturbations. JWST UV LF measurements thus allow the effects of HDE and IHDE on high-redshift galaxy abundances to be examined alongside CMB, BAO, and SN observations.

In this work, we investigate the HDE and IHDE models using CMB observations from Planck and the Atacama Cosmology Telescope (ACT), DESI DR2 BAO measurements, the DES-Dovekie SN sample, and JWST high-redshift UV LF measurements. We aim to use the high-redshift galaxy abundances observed by JWST to provide a complementary test of the HDE and IHDE models, while comparing the predicted high-redshift UV LF of the different models with JWST observations.

This paper is organized as follows. In Sec.~\ref{sec2}, we introduce the HDE and IHDE models, describe the modeling of structure formation and the UV LF, and present the observational datasets used in our analysis. In Sec.~\ref{sec3}, we present the main parameter constraints and discuss the impact of JWST observations on these models. We summarize our conclusions in Sec.~\ref{sec4}.

\section{Methodology and Data}\label{sec2}

\subsection{HDE and IHDE models}\label{sec2.1}

We consider a spatially flat Friedmann--Lemaître--Robertson--Walker universe whose background evolution satisfies
\begin{equation}
3M_{\rm pl}^{2}H^{2}=\rho_{\rm c}+\rho_{\rm de}+\rho_{\rm b}+\rho_{\gamma}+\rho_{\nu},
\end{equation}
where $H$ is the Hubble parameter, $M_{\rm pl}$ denotes the reduced Planck mass, and $\rho_{\rm c}$, $\rho_{\rm de}$, $\rho_{\rm b}$, $\rho_{\gamma}$, and $\rho_{\nu}$ denote the energy densities of CDM, DE, baryons, photons, and neutrinos, respectively.

In the HDE scenario, the DE density is motivated by requiring that an effective field theory in a region of size $L$ should not contain states whose total energy exceeds the mass of a black hole of the same size. Saturating this bound gives
\begin{equation}
\rho_{\rm de}=3c^{2}M_{\rm pl}^{2}L^{-2},
\end{equation}
where $c$ is a dimensionless model parameter and $L$ is the infrared cutoff. Following the HDE construction, $L$ is identified with the future event horizon~\cite{Li:2004rb},
\begin{equation}
L=a(t)\int_{t}^{\infty}\frac{{\rm d}t'}{a(t')}=a\int_{a}^{\infty}\frac{{\rm d}a'}{H(a')a'^{2}}.
\end{equation}

Defining the fractional DE density as $\Omega_{\rm de}=\rho_{\rm de}/(3M_{\rm pl}^{2}H^{2})$, the EoS of HDE is
\begin{equation}
w_{\rm de}=-\frac{1}{3}-\frac{2}{3c}\sqrt{\Omega_{\rm de}}.
\end{equation}
Its asymptotic late-time behavior is governed by $c$: $c<1$ leads to phantom-like evolution and may result in a future big-rip singularity, whereas $c=1$ and $c>1$ correspond to asymptotic de Sitter and quintessence-like evolution, respectively~\cite{Li:2004rb,Zhang:2005hs}.

To explore an interacting extension of HDE, we allow nongravitational energy exchange between DE and CDM. The background continuity equations are
\begin{equation}
\dot{\rho}_{\rm de}+3H(1+w_{\rm de})\rho_{\rm de}=Q,
\end{equation}
and
\begin{equation}
\dot{\rho}_{\rm c}+3H\rho_{\rm c}=-Q,
\end{equation}
where an overdot denotes differentiation with respect to cosmic time. We adopt the commonly studied interaction form
\begin{equation}
Q=\beta H\rho_{\rm de},
\end{equation}
where $\beta$ is a dimensionless coupling parameter. Under this convention, $\beta>0$ corresponds to energy transfer from CDM to DE, whereas $\beta<0$ corresponds to energy transfer in the opposite direction~\cite{Zhang:2007uh,Li:2024qus}.

Combining the evolution of the HDE density with the interacting continuity equation, the dark energy EoS in the IHDE model is
\begin{equation}
w_{\rm de}=-\frac{1}{3}-\frac{2}{3c}\sqrt{\Omega_{\rm de}}+\frac{\beta}{3}.
\end{equation}
The HDE expression is recovered in the noninteracting limit $\beta=0$. The interaction modifies the redshift evolution of the CDM density and consequently affects both the background expansion and the growth of matter perturbations.

\subsection{Structure growth and UV LF modeling}\label{sec2.2}

We follow the numerical modeling and likelihood framework established in Refs.~\cite{Liu:2024yan,Jiang:2024tll,Du:2026qtq}. For the $\Lambda$CDM, HDE, and IHDE models, we recompute the linear matter power spectrum and map it successively to the halo abundance, stellar mass, and UV magnitude.

The halo abundance is calculated using the Sheth--Tormen multiplicity function~\cite{Sheth:1999su}
\begin{equation}
f_{\rm ST}(\nu)=A\sqrt{\frac{2a}{\pi}}\,\nu\exp\left(-\frac{a\nu^{2}}{2}\right)\left[1+\left(a\nu^{2}\right)^{-p}\right],
\end{equation}
where
\begin{equation}
\nu=\frac{\delta_{\rm c}}{\sigma(M_h,z)},\qquad \delta_{\rm c}=1.686.
\end{equation}
The corresponding halo mass function is
\begin{equation}
\frac{{\rm d}n}{{\rm d}M_h}=\frac{\bar{\rho}_{\rm m,0}}{M_h}f_{\rm ST}(\nu)\left|\frac{{\rm d}\ln\sigma^{-1}}{{\rm d}M_h}\right|,
\end{equation}
with
\begin{equation}
M_h=\frac{4\pi}{3}\bar{\rho}_{\rm m,0}R^{3}.
\end{equation}
The parameters $A$, $a$, and $p$ are varied together with the cosmological parameters. The linear density variance is obtained from
\begin{equation}
\sigma^{2}(M_h,z)=\int_{0}^{\infty}\frac{k^{2}\,{\rm d}k}{2\pi^{2}}P(k,z)W^{2}(kR),
\end{equation}
where
\begin{equation}
W(x)=\frac{3\left(\sin x-x\cos x\right)}{x^{3}}.
\end{equation}
Here $P(k,z)$ is the linear matter power spectrum returned by the Boltzmann solver. Internally, we use $k$ in units of $h/{\rm Mpc}$ and $P(k,z)$ in units of $({\rm Mpc}/h)^3$.

We convert halo mass to stellar mass using the double-power-law stellar-to-halo mass relation
\begin{equation}
\frac{M_{*}}{M_{h}}=2N\left[\left(\frac{M_{h}}{M_{\rm c}}\right)^{-\beta_{\rm shmr}}+\left(\frac{M_{h}}{M_{\rm c}}\right)^{\gamma}\right]^{-1}.
\end{equation}
We vary $N$, $M_{\rm c}$, and $\beta_{\rm shmr}$ and fix $\gamma=-0.01$~\cite{CEERSTeam:2023qgy,Scholtz:2023}.

The stellar mass is mapped to the intrinsic UV magnitude through
\begin{equation}
M_{\rm UV}^{\rm int}=-20.5+\frac{\log_{10}(M_{*}/M_{\odot})-a_{0}(z)}{a_{1}(z)}.
\end{equation}
The coefficients $a_0(z)$ and $a_1(z)$ are linearly interpolated over $6\leq z\leq10$ and held fixed at their $z=10$ values for $z>10$~\cite{Stefanon_2021}.

We adopt a UV continuum slope--magnitude relation with coefficients inferred from high-redshift galaxy observations~\cite{Cullen:2023UV},
\begin{equation}
\beta_{\rm UV}=-0.17\,M_{\rm UV}^{\rm int}-5.40.
\end{equation}
The dust attenuation at 1600~\AA\ is then calculated using the Meurer relation~\cite{Meurer:1999jj,Wilkins:2013rlc},
\begin{equation}
A_{1600}=4.43+1.99\,\beta_{\rm UV},
\end{equation}
without imposing a non-negative lower bound on $A_{1600}$. The observed UV magnitude is obtained as
\begin{equation}
M_{\rm UV}=M_{\rm UV}^{\rm int}+A_{1600}.
\end{equation}

The theoretical UV LF is constructed using the deterministic mapping $M_h\rightarrow M_*\rightarrow M_{\rm UV}$ with a bin width of $0.25$ mag.

The likelihood is evaluated in $\log_{10}\Phi$ space,
\begin{equation}
\chi_{\rm JWST}^{2}=\sum_i\frac{\left[\log_{10}\Phi_i^{\rm th}-\log_{10}\Phi_i^{\rm obs}\right]^2}{\sigma_i^2},
\end{equation}
where the upper uncertainty is used when $\Phi_i^{\rm th}\geq\Phi_i^{\rm obs}$ and the lower uncertainty is used otherwise, after transforming the asymmetric errors into $\log_{10}\Phi$ space. Only detections contribute to $\chi_{\rm JWST}^{2}$, while upper limits are used to determine the effective UV magnitude range at each redshift.

\begin{table}[t]
\centering
\footnotesize
\renewcommand{\arraystretch}{1.4}
\setlength{\tabcolsep}{18pt}
\caption{Flat priors on the main cosmological parameters constrained in this paper.}
\label{tab1}
\begin{tabular}{cc}
\hline
\hline
\textbf{Parameter} & \textbf{Prior} \\
\hline
$\Omega_{\rm b} h^2$                  & $\mathcal{U}[0.005,\,0.1]$ \\
$\Omega_{\rm c} h^2$                  & $\mathcal{U}[0.001,\,0.99]$ \\
$\tau$                                & $\mathcal{U}[0.01,\,0.8]$ \\
$\ln(10^{10}A_{\rm s})$               & $\mathcal{U}[1.61,\,3.91]$ \\
$H_0~[\mathrm{km\,s^{-1}\,Mpc^{-1}}]$ & $\mathcal{U}[20,\,100]$ \\
$n_{\rm s}$                           & $\mathcal{U}[0.8,\,1.2]$ \\
$c$                                   & $\mathcal{U}[0.01,\,3]$ \\
$\beta$                               & $\mathcal{U}[-3,\,3]$ \\
\hline
\hline
\end{tabular}
\end{table}

\subsection{Cosmological datasets}\label{sec2.3}

The background evolution and linear perturbations of the HDE and IHDE models are calculated using a customized version of \texttt{IDECAMB}~\cite{Li:2023fdk}, which is developed based on the Boltzmann solver \texttt{CAMB}~\cite{Lewis:1999bs}. In our implementation, the dynamical evolution of HDE is incorporated into the background equations, while the interaction between DE and CDM in the IHDE model is consistently treated at both the background and perturbation levels. For the IHDE model, the extended parametrized post-Friedmann framework is adopted to handle the DE perturbations and avoid the large-scale instabilities that may arise in interacting DE scenarios~\cite{Fang:2008sn,Hu:2008zd,Li:2014eha,Zhang:2017ize}. Cosmological parameter estimation is performed using the Bayesian inference framework \texttt{Cobaya}~\cite{Torrado:2020dgo}, with posterior distributions sampled through the Markov chain Monte Carlo (MCMC) method~\cite{Lewis:2002ah,Lewis:2013hha}. The convergence of the MCMC chains is assessed using the Gelman--Rubin criterion, requiring $R-1<0.02$. Finally, the marginalized posterior distributions and confidence intervals are obtained using \texttt{GetDist}~\cite{Lewis:2019xzd}.

The parameter spaces adopted for the $\Lambda$CDM, HDE, and IHDE models are summarized in Table~\ref{tab1}. We utilize the following cosmological observational datasets:

\begin{itemize}

\item \textbf{CMB}. The CMB dataset combines measurements of the temperature anisotropy, the polarization and temperature--polarization cross-spectra, and the ACT CMB lensing power spectrum. The likelihood comprises four components: (i) at small angular scales, $\ell>30$, we use the $C_{\ell}^{TT}$, $C_{\ell}^{TE}$, and $C_{\ell}^{EE}$ spectra from the Planck PR4 \texttt{CamSpec} likelihood~\cite{Planck:2018vyg,Efstathiou:2019mdh,Rosenberg:2022sdy}; (ii) for the large-scale temperature spectrum over $2\leq\ell\leq30$, we adopt the Planck \texttt{Commander} likelihood~\cite{Planck:2018vyg,Planck:2019nip}; (iii) the E-mode polarization spectrum over the same multipole interval is included through the Planck \texttt{SimAll} likelihood~\cite{Planck:2018vyg,Planck:2019nip}; and (iv) the lensing information is taken from the high-precision reconstruction obtained from the joint analysis of Planck NPIPE PR4 and ACT Data Release~6 data~\cite{Carron:2022eyg,ACT:2023dou,ACT:2023kun}.

\item \textbf{DESI}. The BAO measurements from DESI DR2 are derived from several tracer populations, including the bright galaxy sample, luminous red galaxies, emission-line galaxies, quasars, and the Lyman-$\alpha$ forest. The adopted measurements are those collected in Table~IV of Ref.~\cite{DESI:2025zgx}.

\item \textbf{DES-Dovekie}. The DES-Dovekie SN compilation is based on a recalibration and updated analysis of the Dark Energy Survey five-year sample. It consists of 1623 likely type~Ia supernovae (SNe) satisfying $P_{\rm SNe}>0.5$, together with 197 low-redshift SNe, for a total sample of 1820 SNe~\cite{DES:2025sig}.

\item \textbf{JWST}. The UV LF measurements used in this analysis are derived from recent JWST observations spanning the redshift range $7\leq z\leq12$. The compilation includes the $z=7$ and $z=8$ measurements of \citet{Harikane2024}, the $z=9$--$12$ determinations of \citet{Donnan2024}, the COSMOS-Web results of \citet{Casey2023} at $z=11$, the MIRI-based measurements of \citet{Perez-Gonzalez:2023wta} at $z=9$, 11, and 12, and the faint-end constraints of \citet{Chemerynska2026}. All UV LF measurements are consistently converted to a common comoving-volume convention within the likelihood implementation.

\end{itemize}

\begin{table*}[t]
\renewcommand{\arraystretch}{1.4}
\setlength{\tabcolsep}{8pt}
\centering
\footnotesize
\caption{The $1\sigma$ confidence regions of key cosmological parameters obtained from the CMB, DESI, DES-Dovekie, and JWST data for the $\Lambda$CDM, HDE, and IHDE models. Here, $H_0$ is in units of $\mathrm{km\,s^{-1}\,Mpc^{-1}}$.}\label{tab2}
\begin{tabular}{lccccc}
\hline
\hline
\textbf{Data} & \textbf{$H_0$} & \textbf{$S_8$} & \textbf{$\Omega_{\rm m}$} & \textbf{$c$} & \textbf{$\beta$} \\
\hline
\multicolumn{6}{l}{\textbf{$\Lambda$CDM}} \\
CMB+DESI & $68.17\pm0.28$ & $0.8129\pm0.0076$ & $0.3027\pm0.0037$ & -- & -- \\
CMB+DESI+JWST & $68.17\pm0.28$ & $0.8203\pm0.0074$ & $0.3028\pm0.0035$ & -- & -- \\
CMB+DESI+DES-Dovekie & $68.06\pm0.26$ & $0.8149\pm0.0071$ & $0.3042\pm0.0034$ & -- & -- \\
CMB+DESI+DES-Dovekie+JWST & $68.05\pm0.27$ & $0.8223\pm0.0075$ & $0.3043\pm0.0035$ & -- & -- \\
\hline
\multicolumn{6}{l}{\textbf{HDE}} \\
CMB+DESI & $74.13^{+1.24}_{-1.41}$ & $0.7857\pm0.0075$ & $0.2552\pm0.0087$ & $0.516\pm0.027$ & -- \\
CMB+DESI+JWST & $73.99^{+1.22}_{-1.47}$ & $0.7932\pm0.0077$ & $0.2560\pm0.0089$ & $0.520\pm0.028$ & -- \\
CMB+DESI+DES-Dovekie & $67.79\pm0.54$ & $0.7893\pm0.0081$ & $0.3002\pm0.0049$ & $0.688\pm0.022$ & -- \\
CMB+DESI+DES-Dovekie+JWST & $67.76\pm0.53$ & $0.7989\pm0.0080$ & $0.3003\pm0.0048$ & $0.690^{+0.020}_{-0.023}$ & -- \\
\hline
\multicolumn{6}{l}{\textbf{IHDE}} \\
CMB+DESI & $72.84\pm1.23$ & $1.098^{+0.112}_{-0.265}$ & $0.128^{+0.034}_{-0.069}$ & $0.810\pm0.134$ & $0.519^{+0.234}_{-0.093}$ \\
CMB+DESI+JWST & $72.83^{+1.15}_{-1.41}$ & $0.949^{+0.075}_{-0.099}$ & $0.170^{+0.032}_{-0.044}$ & $0.723\pm0.099$ & $0.380^{+0.176}_{-0.100}$ \\
CMB+DESI+DES-Dovekie & $67.98\pm0.53$ & $1.277\pm0.183$ & $0.094^{+0.013}_{-0.042}$ & $1.270^{+0.146}_{-0.096}$ & $0.650^{+0.104}_{-0.034}$ \\
CMB+DESI+DES-Dovekie+JWST & $67.88\pm0.52$ & $1.029^{+0.065}_{-0.086}$ & $0.152^{+0.024}_{-0.031}$ & $1.091\pm0.095$ & $0.498^{+0.090}_{-0.061}$ \\
\hline
\hline
\end{tabular}
\end{table*}

\begin{figure*}[t]
\centering
\includegraphics[width=0.9\textwidth]{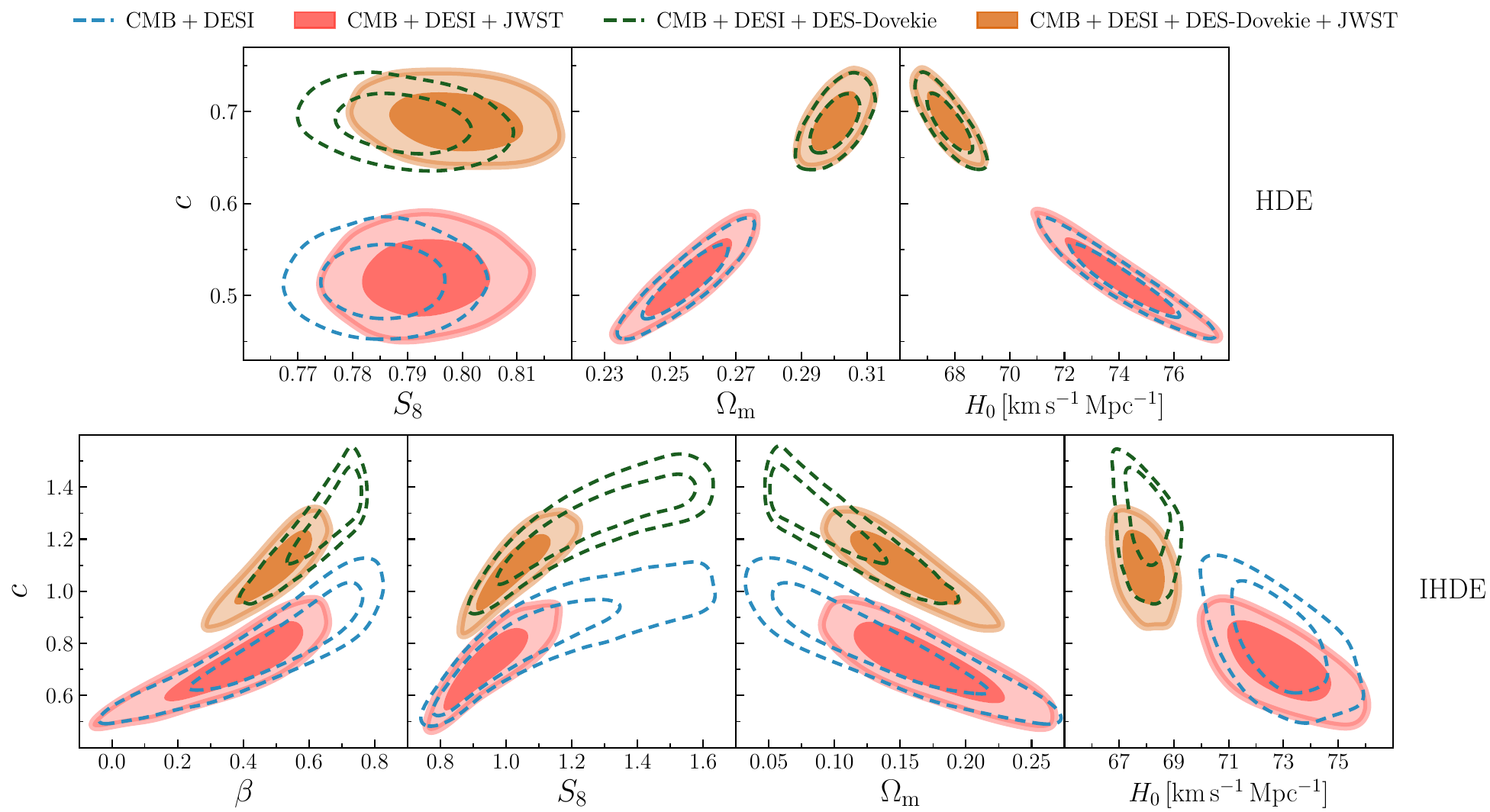}
\caption{Constraints on the cosmological parameters using the CMB+DESI, CMB+DESI+JWST, CMB+DESI+DES-Dovekie, and CMB+DESI+DES-Dovekie+JWST data for the HDE (upper) and IHDE (lower) models.}
\label{fig1}
\end{figure*}

\begin{figure*}[t]
\centering
\includegraphics[width=\textwidth]{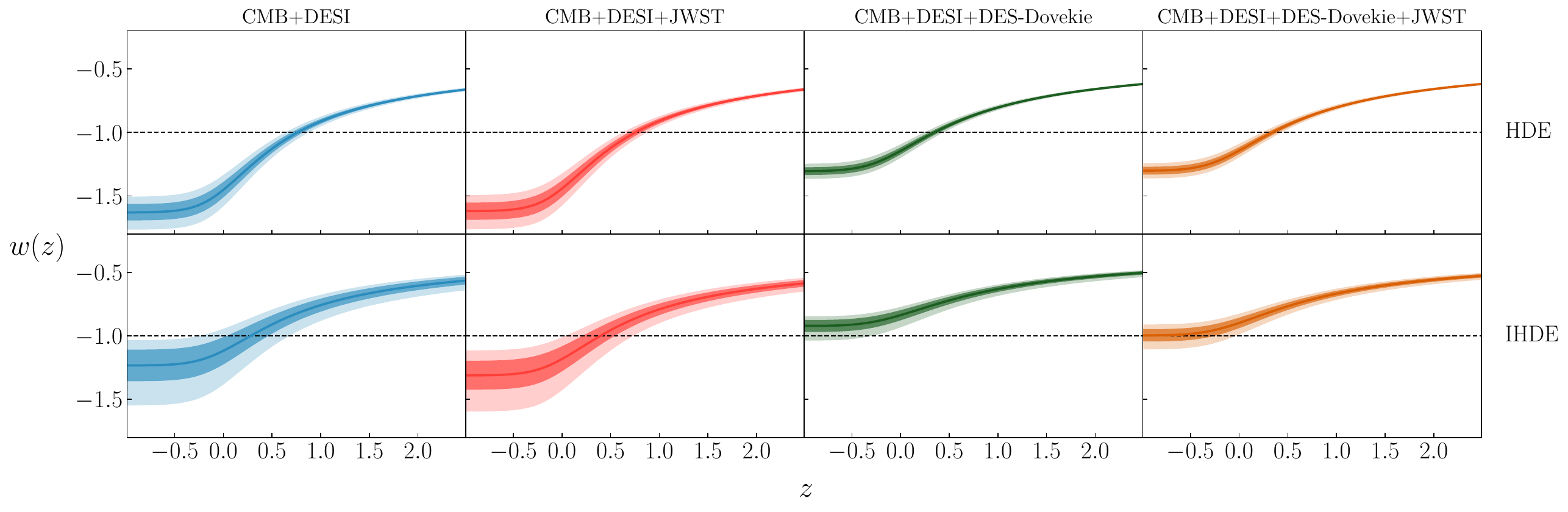}
\caption{Reconstructed evolution histories of $w(z)$ at the $1\sigma$ and $2\sigma$ confidence levels in the HDE (upper) and IHDE (lower) models, constrained by the CMB+DESI, CMB+DESI+JWST, CMB+DESI+DES-Dovekie, and CMB+DESI+DES-Dovekie+JWST data. The black dashed line denotes the phantom divide $w_{\rm de}=-1$.}
\label{fig2}
\end{figure*}

\begin{table*}[t]
\renewcommand{\arraystretch}{1.4}
\setlength{\tabcolsep}{3pt} 
\centering
\footnotesize
\caption{Constraints on the astrophysical parameters from the data combinations including the JWST data. The characteristic halo mass $M_{\rm c}$ is expressed in units of $M_\odot/h$.}\label{tab3}
\begin{tabular}{lcccccc}
\hline\hline
\textbf{Data} & \textbf{$A$} & \textbf{$a$} & \textbf{$p$} & \textbf{$\log_{10} M_{\rm c}$} & \textbf{$N$} & \textbf{$\beta_{\rm shmr}$} \\
\hline
\multicolumn{7}{l}{\textbf{$\Lambda$CDM}} \\
 CMB+DESI+JWST & $0.3292 \pm 0.0020$ & $0.7678 \pm 0.0030$ & $0.260 \pm 0.020$ & $10.773^{+0.053}_{-0.061}$ & $0.0201^{+0.0015}_{-0.0019}$ & $1.99 \pm 0.13$ \\
 CMB+DESI+DES-Dovekie+JWST & $0.3292 \pm 0.0020$ & $0.7680 \pm 0.0030$ & $0.260 \pm 0.021$ & $10.767^{+0.049}_{-0.062}$ & $0.0200^{+0.0015}_{-0.0019}$ & $2.01 \pm 0.13$ \\
\hline
\multicolumn{7}{l}{\textbf{HDE}} \\
 CMB+DESI+JWST & $0.3292 \pm 0.0020$ & $0.7678 \pm 0.0030$ & $0.260 \pm 0.020$ & $10.809^{+0.055}_{-0.066}$ & $0.0201^{+0.0016}_{-0.0019}$ & $2.00 \pm 0.13$ \\
 CMB+DESI+DES-Dovekie+JWST & $0.3292 \pm 0.0020$ & $0.7678 \pm 0.0030$ & $0.260 \pm 0.020$ & $10.789^{+0.057}_{-0.068}$ & $0.0202^{+0.0017}_{-0.0020}$ & $1.98 \pm 0.14$ \\
\hline
\multicolumn{7}{l}{\textbf{IHDE}} \\
 CMB+DESI+JWST & $0.3292 \pm 0.0020$ & $0.7678 \pm 0.0030$ & $0.260 \pm 0.020$ & $10.610 \pm 0.120$ & $0.0311^{+0.0053}_{-0.0081}$ & $1.99 \pm 0.13$ \\
 CMB+DESI+DES-Dovekie+JWST & $0.3292 \pm 0.0020$ & $0.7678 \pm 0.0030$ & $0.259 \pm 0.020$ & $10.502 \pm 0.099$ & $0.0372^{+0.0055}_{-0.0074}$ & $1.98 \pm 0.13$ \\
\hline\hline
\end{tabular}
\end{table*}

\begin{figure*}[t]
\centering
\includegraphics[width=0.95\textwidth]{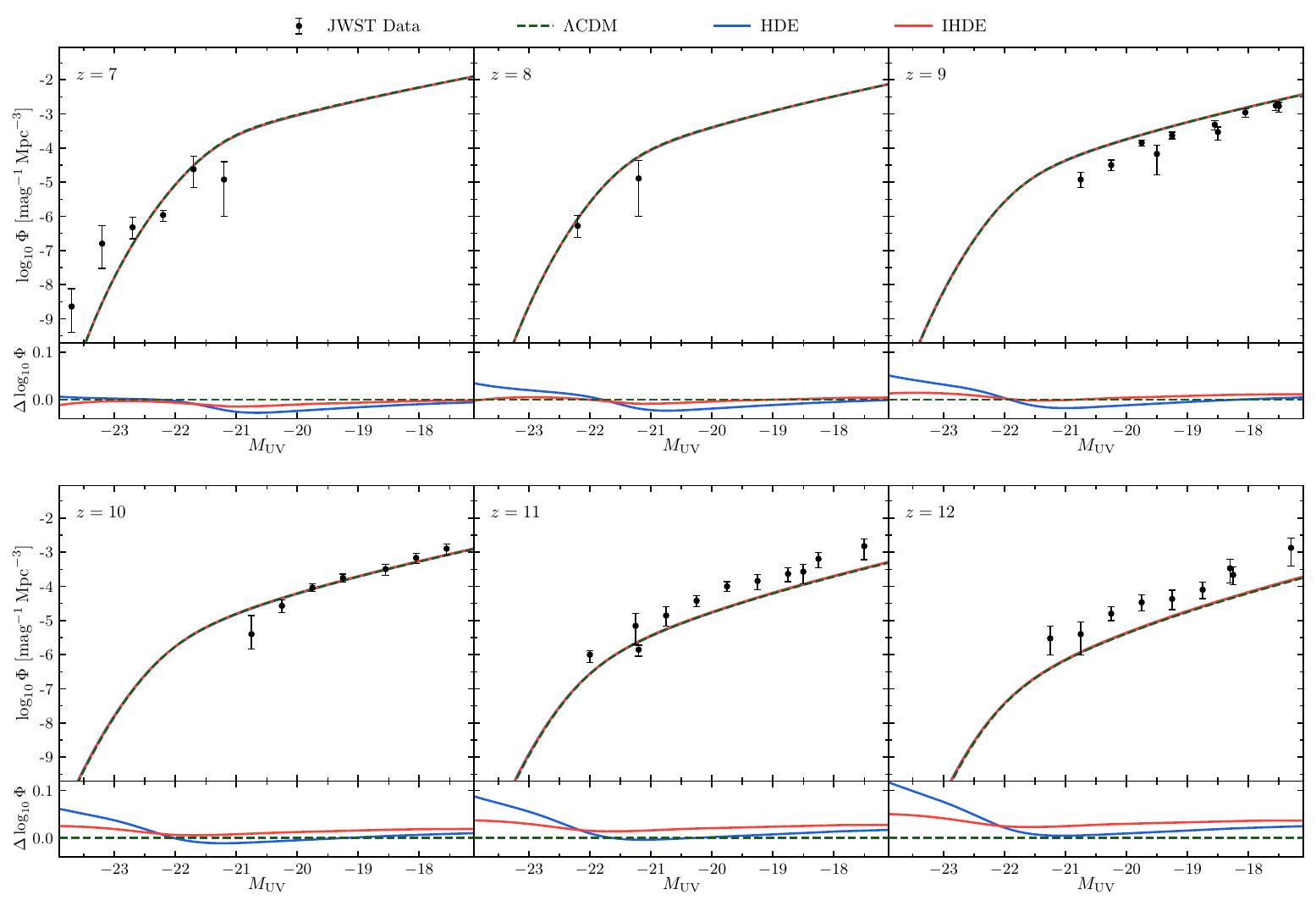}
\caption{Theoretical predictions for the UV LF $\log_{10}\Phi$ in the $\Lambda$CDM, HDE, and IHDE models over the redshift range $7\leq z\leq12$, obtained using the CMB+DESI+DES-Dovekie+JWST data and compared with the JWST measurements. For each redshift, the main panel shows the predicted UV LF, while the lower residual panel shows $\Delta\log_{10}\Phi$ relative to the $\Lambda$CDM model.}
\label{fig3}
\end{figure*}

\section{Results and Discussion}\label{sec3}

In this section, we present the main cosmological constraints, as summarized in Table~\ref{tab2} and Fig.~\ref{fig1}. We also present the reconstructed EoS in Fig.~\ref{fig2}. The constraints on the astrophysical parameters and the UV LF results are presented in Table~\ref{tab3} and Fig.~\ref{fig3}, respectively.

As shown in Table~\ref{tab2}, in the $\Lambda$CDM model, adding the JWST data leaves $H_0$ and $\Omega_{\rm m}$ nearly unchanged and produces only a modest upward shift in $S_8$. For the HDE model, the inclusion of the JWST data likewise has only a limited impact on the constraints on $H_0$, $\Omega_{\rm m}$, and the holographic parameter $c$, with the posterior distributions in the upper panel of Fig.~\ref{fig1} showing that its main effect is a mild upward shift in $S_8$. By comparison, the DES-Dovekie data have a much more pronounced impact on the HDE parameters, shifting $H_0$ to substantially lower values while increasing $\Omega_{\rm m}$ and $c$.

For the IHDE model, the constraints generally yield higher $S_8$ and lower $\Omega_{\rm m}$. For example, the CMB+DESI+DES-Dovekie data give $S_8=1.277\pm0.183$, $\Omega_{\rm m}=0.094^{+0.013}_{-0.042}$, and $\beta=0.650^{+0.104}_{-0.034}$. A positive $\beta$ indicates energy transfer from CDM to DE, further reducing the CDM density relative to the noninteracting case and yielding a lower present-day matter density parameter of approximately $0.1$. At the perturbation level, the interaction in IHDE also modifies the evolution of matter perturbations, corresponding to enhanced matter clustering in this parameter region and resulting in a higher $S_8$.

When JWST data are added, as shown in the lower panel of Fig.~\ref{fig1}, $c$ and $\beta$ for CMB+DESI+DES-Dovekie shift from $1.270^{+0.146}_{-0.096}$ and $0.650^{+0.104}_{-0.034}$ to $1.091\pm0.095$ and $0.498^{+0.090}_{-0.061}$, respectively, while $\Omega_{\rm m}$ increases to $0.152^{+0.024}_{-0.031}$ and $S_8$ decreases to $1.029^{+0.065}_{-0.086}$. These results indicate that JWST substantially reduces posterior support for the stronger interaction region characterized by low $\Omega_{\rm m}$ and high $S_8$, while tightening the constraint on $c$. For CMB+DESI, a similar trend is also observed when JWST data are added. One possible explanation is that the observed high-redshift galaxy abundances require a sufficient population of massive dark matter halos, whereas some IHDE parameter regions with stronger interactions and lower $\Omega_{\rm m}$ may predict too few halos, thereby becoming disfavored by JWST and shifting the posterior toward weaker interactions, higher $\Omega_{\rm m}$, and lower $S_8$. The panels of Fig.~\ref{fig1} show substantial posterior shifts in HDE and IHDE when DES-Dovekie is added to CMB+DESI, with limited overlap between the credible regions in several two-dimensional parameter planes.

To further investigate the dynamical evolution of DE in the HDE and IHDE models, we reconstruct the EoS $w(z)$, as shown in Fig.~\ref{fig2}. In the HDE model, all data combinations favor $c<1$, and the median $w(z)$ curves cross the phantom divide $w=-1$, indicating a preference for phantom evolution and the associated future Big Rip behavior characteristic of HDE. After introducing the dark sector interaction, $w(z)$ in the IHDE model shifts toward larger values, substantially weakening the phantom behavior, most notably for the combinations including the DES-Dovekie data. This behavior can be understood directly from the IHDE EoS, in which a positive $\beta$ contributes positively to $w_{\rm de}$, while a larger holographic parameter $c$ suppresses the negative holographic contribution. The dark sector interaction can therefore partially counteract the phantom evolution of HDE and reduce its tendency toward a Big Rip future. Furthermore, for both the CMB+DESI and CMB+DESI+DES-Dovekie baselines, adding the JWST data shifts both $c$ and $\beta$ toward lower values, and both changes drive $w(z)$ toward more negative values, consistently explaining the mildly enhanced phantom behavior seen in Fig.~\ref{fig2}. In particular, although the inclusion of the JWST data shifts $w(z)$ toward more negative values for the combinations including DES-Dovekie, the phantom behavior in IHDE remains substantially weaker than in the corresponding HDE case.

Table~\ref{tab3} summarizes the constraints on the astrophysical parameters from the data combinations including the JWST data. Across the three cosmological models, the constraints on $A$, $a$, $p$, and $\beta_{\rm shmr}$ remain stable, while the galaxy--halo relation parameters obtained in the HDE model are broadly consistent with those in the $\Lambda$CDM model. The main differences appear in the IHDE model, which favors a lower characteristic halo mass and a higher normalization. For example, for the CMB+DESI+DES-Dovekie+JWST data, $\log_{10} M_{\rm c}$ decreases from $10.767^{+0.049}_{-0.062}$ in the $\Lambda$CDM model to $10.502\pm0.099$ in the IHDE model, while $N$ increases from $0.0200^{+0.0015}_{-0.0019}$ to $0.0372^{+0.0055}_{-0.0074}$. This indicates that the altered structure-formation history in the IHDE model is accompanied by a simultaneous adjustment of the galaxy--halo mapping.

To further examine the predictions of the three cosmological models for high-redshift galaxy abundances, we present their UV LFs for the CMB+DESI+DES-Dovekie+JWST data, together with their residuals relative to $\Lambda$CDM, as shown in Fig.~\ref{fig3}. At $z=7$--$9$, the predicted UV LFs are nearly indistinguishable in the main panels, with differences among the models generally limited to a few hundredths of a dex\footnote{Here, dex denotes a difference on a base-10 logarithmic scale.}. These differences become progressively more pronounced toward higher redshift and are concentrated mainly at the bright end of the UV LF. In particular, at $z=10$--$12$, the HDE model produces the largest positive deviation from $\Lambda$CDM at bright magnitudes, reaching approximately $0.1$ dex at the brightest end for $z=11$ and $z=12$. Toward fainter magnitudes, this deviation rapidly decreases toward zero and becomes slightly negative over part of the magnitude range. By comparison, the IHDE model exhibits smaller and smoother deviations from $\Lambda$CDM across the magnitude range. Although the different DE models produce systematic differences at the bright end at high redshift, several JWST measurements at $z=11$ and $z=12$ still lie above the predictions of all three models. Overall, although the differences in the UV LF among the models remain smaller than the observational uncertainties of individual data points, their cumulative effects can still influence the goodness of fit. For the CMB+DESI+DES-Dovekie+JWST data, $\chi^2_{\rm JWST}$ decreases by $6.91$ for HDE and $7.12$ for IHDE compared with $\Lambda$CDM. These results indicate that both HDE and IHDE provide better fits to the JWST data, partly because their modified high-redshift UV LFs bring the predicted galaxy abundances closer to some of the observations, particularly at the bright end.

\section{Conclusions}\label{sec4}

In this work, we combine CMB, BAO, and SN observations with JWST high-redshift UV LF measurements to investigate the impact of JWST data on the $\Lambda$CDM, HDE, and IHDE models.

We find that JWST has only a limited impact on the main cosmological parameters of $\Lambda$CDM and HDE, producing a mild upward shift in $S_8$. By contrast, IHDE exhibits a more pronounced response. For CMB+DESI, adding JWST shifts $c$ from $0.810\pm0.134$ to $0.723\pm0.099$ and $\beta$ from $0.519^{+0.234}_{-0.093}$ to $0.380^{+0.176}_{-0.100}$, accompanied by an increase in $\Omega_{\rm m}$ and a decrease in $S_8$. Similar trends are found when JWST is added to CMB+DESI+DES-Dovekie. These results indicate that JWST data reduce the posterior support for the stronger interaction region characterized by high $S_8$ and low $\Omega_{\rm m}$. The corresponding EoS of DE shifts toward more negative values, although the phantom behavior in IHDE remains generally weaker than in HDE.

The predicted UV LFs are broadly similar at $z=7$--$9$, while model differences become more pronounced at higher redshifts, particularly at the bright end. Both HDE and IHDE improve the fit to JWST data relative to $\Lambda$CDM, reducing $\chi^2_{\rm JWST}$ by approximately $7$. More precise high-redshift UV LF measurements, together with more flexible galaxy-halo modeling, will help further test HDE and IHDE.

\section*{Acknowledgments}
We thank Jun-Qian Jiang, Zhao-Yu Li, and Hui Li for their helpful discussions. This work was supported by the National Natural Science Foundation of China (Grants Nos. 12533001, 12575049, and 12473001), the National SKA Program of China (Grants Nos. 2022SKA0110200 and 2022SKA0110203), the China Manned Space Program (Grant No. CMS-CSST-2025-A02), and the National 111 Project (Grant No. B16009).

\bibliography{main} 

\end{document}